# Semimetallic molecular hydrogen at pressure above 350 GPa

M. I. Eremets*, A. P. Drozdov, P. P. Kong, H. Wang

Max-Planck-Institut fur Chemie, Hahn-Meitner Weg 1, 55128 Mainz, Germany

**According to the theoretical predictions, insulating molecular hydrogen dissociates and transforms to an atomic metal at pressures P~370-500 GPa[1-3]. In another scenario, the metallization first occurs in the 250-500 GPa pressure range in molecular hydrogen through overlapping of electronic bands[4-7].The calculations are not accurate enough to predict which option is realized. Here we show that at a pressure of ~360 GPa and temperatures <200 K the hydrogen starts to conduct, and that temperature dependence of the electrical conductivity is typical of a semimetal. The conductivity, measured up to 440 GPa, increases strongly with pressure. Raman spectra, measured up to 480 GPa, indicate that hydrogen remains a molecular solid at pressures up to 440 GPa, while at higher pressures the Raman signal vanishes, likely indicating further transformation to a good molecular metal or to an atomic state.**

Achieving a metallic state of hydrogen, predicted to occur at high pressure, is one of the most attractive goals in condense matter physics and remains a long-standing challenge both for theory and experiment. In 1935 Wigner and Huntington[1] proposed that any lattice built of hydrogen atoms (protons) should display metallic properties – similar to the alkali metals. However, a metallic state can be stabilized only at very high pressures ~370-500 GPa [2,3,8]. Besides the ultimate simplicity, atomic metallic hydrogen is attractive because of the predicted very high critical temperature for superconductivity[9]. Recently, experimental evidence on the transformation of hydrogen to the atomic state at 495 GPa was reported[10]. This work was met with strong criticism[11]: in particular, the pressure is likely significantly (>100 GPa) overestimated, and the observed enhanced reflectance could be related to a transformation observed in earlier work at ~360 GPa[12]. There is another possibility for transformation to a metallic state: the band gap of the crystalline molecular hydrogen can decrease with pressure and eventually close prior the dissociation of molecules and transformation to the atomic state. This path to metallization is considered in many recent theoretical estimates[4-7]. It also requires very high pressures of ~250-500 GPa.

The calculations and prediction of metallization rely on the knowledge of the structure, however, only the structure of phase I (Fig. 1) was determined as $P6_3/mmc$ at P=5.4 GPa[13]. The structure of phase III (the subject of the present study) still remains unidentified[14,15]. *Ab initio* structural predictions suggest that *C2/c* structure is the most likely candidate for phase III at P>200 GPa[2]. This structure generally agrees with the Raman and infrared data available in the 150-300 GPa[16] range while the quantitative description of the infrared spectra is not satisfactory[17]. The DFT-based methods which are used in the crystal structure search are not suitable for calculations of the band-gap, where the gap is strongly underestimated. The GW (Green's function approximation) is better, and gives the gap closure in the C2/c structure at ~360 GPa[18]. Diffusion quantum Monte Carlo (DMC) calculations give similar pressure of P ~335 GPa for the closure[5]. In these and the majority of other calculations the *indirect* (thermal) gap closes first. This means that the closure is difficult to detect optically because the absorption associated with indirect, phonon-assisted transitions is low.

The typical absorption coefficient is $\alpha$~10-$10^2$ $cm^{-1}$, while a sample in DACs is ~$10^{-4}$ cm thick. Only the direct gap can be measured [16,19] where the absorption coefficients $\alpha>10^4$ $cm^{-1}$.

Electrical measurements are indispensable for the detection of the closure of the indirect gap and the metallization. As soon as the bands in different points of the reciprocal space overlap, electrons in the conduction band and holes in the valence band appear and conductivity can be easily detected. At the beginning of the band overlap, the density of electrons and holes is low and the conductivity poor. This state is typically called "a semimetal" but it is *a metal* because electrons and holes are free and the electrical conductivity is finite down to the lowest temperatures.

In our experiment, we pressurized hydrogen at low temperatures below 200 K, *i. e.* in phase III, and measured electrical conductivity and Raman spectra. The Raman spectra are in agreement with previously published data[16] and we extended the measurements from ~320 GPa[16] to 480 GPa (Fig. 2). Hydrogen remains in the same molecular phase III as the Raman peaks shifts smoothly with pressure (Fig. 2d,e). In contrast to the steady pressure dependence of the energy of the peaks, intensities of the Raman peaks for both the phonon and vibrons drop sharply at 330-360 GPa (Fig. 2f and Refs[12,20]). At higher pressures, up to 400 GPa (Fig. 2f), the intensity of the Raman peaks is nearly constant, and then it decreases, and the Raman peaks disappear above 440 GPa (Fig. 2a,b).

In the electrical measurements, hydrogen is not conducting in the beginning, but visually the sample darkens, and reflects well at ~350 GPa (Ref.[20], SI Fig. 2), the reflection increases with pressure. At 352 GPa and 190 K (at 365 GPa and 184 K in another run) hydrogen starts to conduct. Note that hydrogen is also conductive at P>~280 GPa in phase V at T>200 K (SI Fig. 3e, Fig. 1)[21]. The resistivity $\rho$ of hydrogen in phase III strongly decreases with pressure (Fig. 3c). The temperature dependence $\rho(T)$ indicates a typical metallic behaviour (decrease of $\rho$ at cooling) starting from ~200 K. However, the behaviour of $\rho(T)$ changes and it increases upon further cooling (Fig. 3). This upturn of $\rho$ cannot be attributed to a semiconducting state because the activation energy is very small: it is $E_a$~20 meV at 360 GPa and drops below 6 meV (SI Fig. 3c) with pressure. We emphasize that the measured $\rho(T)$ relates to the hydrogen sample as the same $\rho(T)$ was obtained at different environments of the hydrogen sample including very different materials of the gasket and arrangement of the electrodes. For instance, $\rho(T)$ in SI Fig. 3e was obtained with electrodes deposited on opposite anvils and therefore a possible contribution from conductivity of diamond at these high pressures is excluded.

The electrical measurements are consistent with transformation of hydrogen to a metal through an overlap of distinct energy bands at P>~360 GPa. The low concentration of carriers explains the high electrical resistivity of hydrogen and its decrease with pressure as the overlap of electronic bands increases. In this semimetallic state, the Fermi energy can be small (only ~100 K for typical semimetals[22]) and, according to the Fermi-Dirac distribution, the numbers of electrons and holes decreases with decreasing of temperature (but remains finite at zero temperature). This notably temperature-dependent carrier density *n* naturally explains the increase of the resistivity as temperature falls, and its saturation (Fig. 3c, SI Fig. 3). Our data agree with the theoretical calculations which show appearance of DC conductivity in the C2/c structure at pressure around 335 GPa[5]. We did not observe superconductivity likely due to low carrier density. There are only a few examples of superconductors with very small Fermi surface such as semimetallic Bi[22] or oxygen[23]..

It is useful to compare the metallization in hydrogen with other materials. The value of the resistance and its temperature dependence measured in hydrogen is typical for semimetals, for instance, for bismuth, an archetype semimetal[22]. We found a striking similarity of hydrogen with xenon (Fig. 3d, SI Fig. 4). The metallization of xenon at ~130 GPa[24] through closing of the indirect gap has been well established[25]. The resistivity of xenon $\rho$~$10^{-4}$ Ohm·m at the pressures close to its metallization is close to hydrogen resistivity at 440 GPa (SI Fig. 4b). For further comparison, we extended electrical measurements of xenon in the present work to 250 GPa (Fig. 3d, SI Fig. 4). The resistivity of xenon drops with pressure and saturates only at P~200 GPa, *i. e.* ~1.5 times higher than the pressure of metallization. This may suggest that hydrogen will have a similar value of the resistivity at P~500 GPa.

Oxygen, iodine and bromine[23] metallize in a similar way through the closing of indirect gaps[26]. In oxygen, the Raman signal from the semimetallic ζ-phase is well defined[27], the value of resistivity and the temperature dependence is similar to hydrogen: it increases as temperature falls[28] (iodine and bromide behave similarly[23]). Oxygen is already superconducting at 120 GPa[23] when the resistivity in a normal state is high: ~1x10$^{-5}$ Ohm·m. At similar resistivity, superconductivity in hydrogen can be expected at pressures ~500 GPa (Fig. 3d).

Our Raman data evidence that the semimetallic hydrogen remains in the molecular form (Fig. 2). The persistence of the phase III to pressures up to 440 GPa is in agreement with recent advanced DMC calculations[3,5,29,30] which show stability of the *C*2/*c* structure in the phase III over the wide pressure range close to the transition to the atomic Cs-IV structure occurring at 450-500 GPa[3,5]. On the contrary, the existence of the phonon and vibron Raman bands contradict the prediction of the atomic state at 378 GPa[8].

The peak in the Raman intensity at pressures of ~310-320 GPa (Fig. 3e), is likely a resonance of the Raman scattering with the HeNe laser excitation (1.92 eV). The pressure corresponding to the resonance is in a reasonable agreement with the extrapolation of the resonance peaks observed in Ref. [16] at other laser wavelengths. The energy of the resonance does not help however to determine the indirect or direct gaps because the vertical transitions of the resonance are associated with critical points of the Brillouin zone related to the maximum of polarizability. The indirect or direct gaps have significantly lower energies (for Ge[31], for instance, the resonance peak is at $E_{res}$ = 2.22 eV, while $E_{indr}$ ~0.74 eV, $E_{dir}$ ~0.9 eV).

After the indirect band closure at ~360 GPa, the persistence of the Raman signal is in agreement with the weak absorption and the low screening of the light due to the small amount carriers. Further drop of the intensity at P>400 GPa may be associated with the closure of the direct gap, or a transformation to another, metallic molecular phase. Finally, the vanishing of the Raman signal at P~450 GPa might be connected with a transformation to atomic hydrogen[3].

Our combined Raman and electrical measurements allow us to specify the domain of the metallic hydrogen on the phase diagram (Fig. 1): molecular hydrogen transforms to a semimetal state at ~350 GPa at low temperature in phase III.

Acknowledgements. Support provided by the European Research Council under Advanced Grant 267777 is acknowledged. The authors appreciate Th. Timusk, V. Kresin, L. Boeri, F. Balakirev, Sh. Mozaffari and D. Graf for helpful discussions and comments. ME is thankful the Max Planck community for the invaluable support, and U. Pöschl for the constant encouragement.

Author contributions. Author contributions: All authors equally participated in the experiments. M.E. designed the study and wrote the manuscript together with A. D.

**Methods**

Combined electrical and Raman measurements were performed in a diamond anvil cell (DAC). We used bevelled anvils with culets of diameter 18-20 µm. We polished the anvils by ourselves with the traditional method of using a cast disk charged with diamond powder. After polishing, the anvils were etched in plasma to remove polishing defects. Hydrogen was loaded by condensing it at $T < 20$ K in a cavity around the anvils and clamping it in liquid or solid state. The thickness of the sample was 1 - 2 µm when loaded in liquid state in the hole, or ~0.1 µm when it was clamped in solid state on the surface of the gasket. In another method, a diamond anvil cell (DAC) was placed in a gas loader at pressure about 100 MPa and clamped. After pressurizing to about 100 GPa, the cell was placed into a cryostat and further pressurized at low temperatures, typically at about 100 K.

A schematic arrangement for electrical measurements is shown in Fig. 3. Electrodes (tantalum covered with gold) were sputtered directly on the surface of a diamond anvil (Fig. 3a). In one experiment, all four electrodes touched the sample (SI Fig. 1). In the two another runs, two or three electrodes touched the sample, and the resistance of the electrodes (~100 Ohms) was added to the measurements. We did not use any protection layer (such as $Al_2O_3$) against hydrogen because the experiments were done at low temperatures. The electrodes were separated from the metallic gasket by an insulation layer made of different materials, for instance, $CaF_2$ mixed with epoxy. In all experiments the measurable conductivity was recorded only between electrodes that touches the hydrogen sample; no conductivity was observed between electrodes that did not touch the hydrogen sample. Nevertheless, since in principle there could be other sources of conductance in the experiment other than the sample, we carefully examined other possibilities. The contact of the electrodes with the metallic gasket was monitored and no shortage was observed. The insulating layer between the metallic gasket and the diamond anvil is not conducting because we reproduced the same results in runs with different insulating materials: $CaSO_4$, $CaF_2$ mixed with epoxy, NaCl. MgO mixed with epoxy was used in Ref. [1] (Ref. 1 Methods). Usage of different gasket materials also rules out a possible reaction of hydrogen with the gasket. A contribution to the measured conductivity from the diamond is unlikely or very small because the same result (a temperature dependence of the resistance and a similar value of the resistance (Fig. 3)) was obtained in Ref. [1] (Ref. 1 Methods) (Fig. 9a), where electrodes were sputtered on the opposite anvils. A strong argument that conductivity is associated only with hydrogen is the drop of resistance at the transition from phase V to phase III (SI Fig. 3e), which is consistent with the change in the Raman spectra (SI Fig. 5).

We measured the Raman spectra of hydrogen through synthetic pure diamond anvils (type IIa), which have very low intrinsic luminescence even at the highest pressures (Figs 2, SI Figs 1,5). At a pressure of ~300 GPa, we reduced the intensity of the HeNe laser by a factor of 5-10 to <1 mW to the minimum level where satisfactory Raman spectra can be obtained. This precaution was taken to prevent the known failure of diamonds, which happens under an illumination at the highest pressures.

The pressure was determined from the shift of the Raman spectra of the stressed diamonds anvils originally established in Ref.[2](Ref. 2 Methods) up to 200 GPa and then extended to ~400 GPa[3] (Ref. 3 Methods). The step at the high frequency edge of the diamond spectra (Fig. 2a,c, SI Fig. 1) was well resolved and can be used for a reliable determination of the maximum pressure achieved as 480 GPa. The step was clearly observed to the highest pressures and is assigned to 480 GPa (Fig. 2a,c) according to Akahama[3], or to 500 GPa in our updated scale[2]. These scales were calibrated against the equation of state (EOS) of gold measured at 400 GPa, and then extrapolated to higher pressures. For consistency we use the scale from Ref [3] in this work.

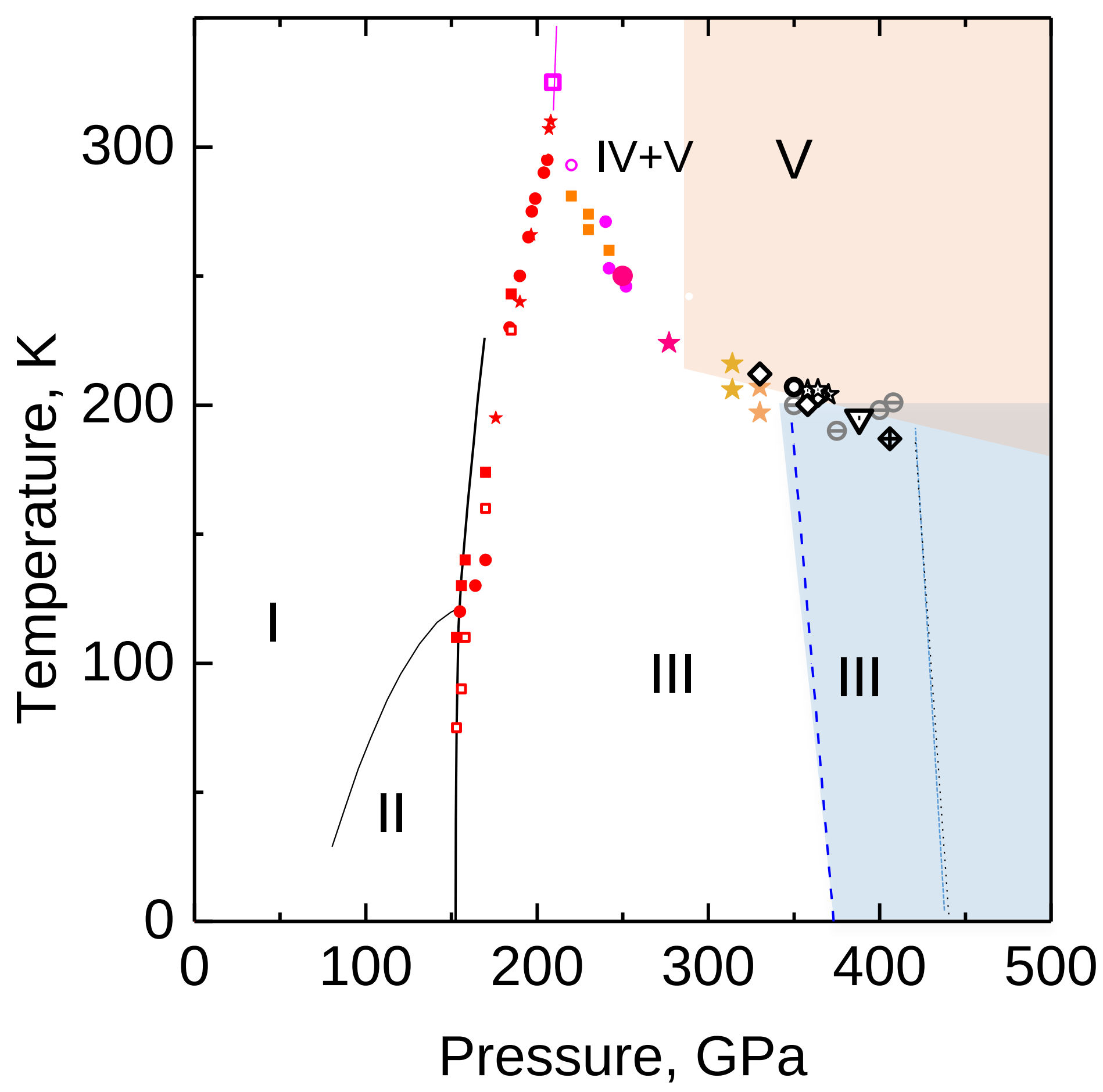

Fig. 1. Low-temperature part of the phase diagram of hydrogen. Domains of different phases are indicated in the Roman numerals (see Ref. 13) for details. The black experimental points are obtained in the present work. The other points are from Ref. 13. Different colors and symbols indicate different experimental runs. The shaded blue region indicates the domain of the molecular but metallic (semimetallic) hydrogen in phase III, the shaded yellow region indicate the domain of the phase V. A dashed line at 440 GPa indicates a possible boundary with the next nonmolecular phase (or good metallic molecular state) as signatures of hydrogen molecules in the Raman spectra disappeared at higher pressures.

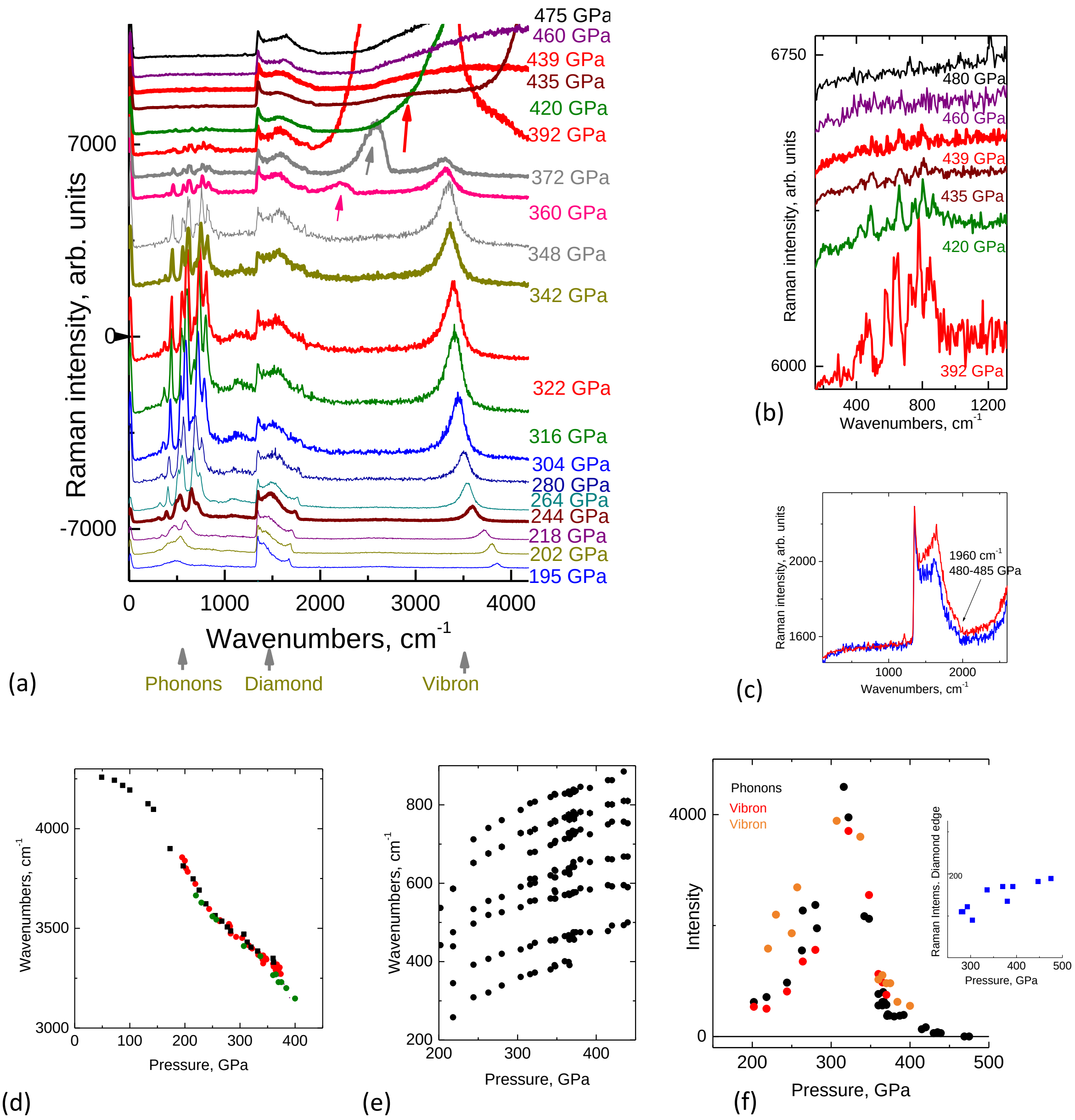

Fig. 2. Raman scattering from hydrogen taken at T~100 K.

(a) Raman spectra from hydrogen shifted with respect to each other to follow the transformation of the spectra. The ranges of characteristic excitations (phonon and vibron, and the signal from the stressed diamonds) are shown by arrows at the bottom. At a pressure of ~350 GPa, a luminescence appeared, its peak is indicated by arrow. With pressure, it shifts to higher wavenumbers (meaning that the energy of the luminescence peak decreases) and its intensity becomes huge and it covers the vibron peak. It is associated with the stressed diamond, not the hydrogen sample (SI Fig. 6 for further details), in particular, this luminescence appeared at different pressures of, for instance, 310 GPa, and 210 GPa in other runs.

(b) The phonon spectra at the highest pressures showing gradual vanishing of the Raman signal and its disappearing at a pressure between 439-460 GPa.

(c) Raman spectra from the stressed diamond at the highest pressure taken at two different points of the sample. The arrow indicates a step – from its position the pressure is determined (see Methods for details).

(d) The pressure dependence of the vibron peak. A kink on the pressure dependence of vibron at ~240 GPa is likely related to the phase transformation from the hexagonal $P6_122$ structure to the $C2/c$ structure of molecular hydrogen (Phys. Rev. B **94** 134101). Symbols in different colors relate to different samples.

(e) The pressure dependence of phonons. The smooth dependence indicates that there are no major structural transitions.

(f) The intensity of the phonon Raman signal (plotted for strongest peak (it is at ~740 cm$^{-1}$ at 322 GPa), other peaks behave similarly with pressure), and the vibron peak (for two runs indicated by different colors) plotted versus pressure. The intensity drops at pressures above ~320 GPa. Even though it is much weaker at pressures above ~330 GPa, the phonon spectra can be tracked up to 440 GPa. The changes in the intensity are related to the sample and not to an increase of absorption by the diamond anvil, as the Raman signal from diamond does not change appreciably with pressure (inset in (f)).

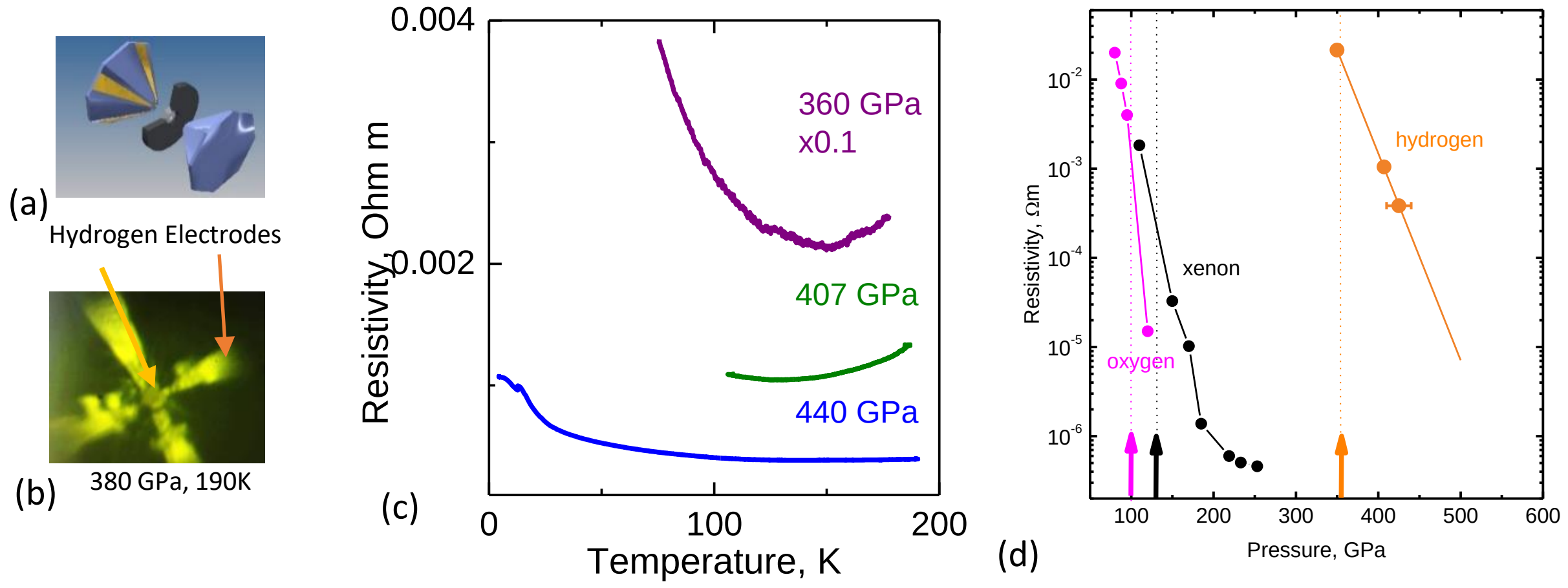

Fig. 3. Electrical resistance of hydrogen at different pressures.

(a) Arrangement of DAC for electrical measurements. Four electrodes (tantalum covered by gold) are sputtered on the upper anvil. Metallic gasket with an insulating layer (indicated with white color) is between the two anvils. Anvils and gasket are separated for clarity.

(b) Photograph of the hydrogen sample (a reflective spot in the center) together with four attached electrodes at 380 GPa and 190 K. The photo was taken in reflection illumination.

(c) Temperature dependence of the resistivity at different pressures.

(d) Pressure dependence of resistivity (the points are taken at 120 K). For comparison, the pressure dependence of resistivity of oxygen and xenon is shown. The arrows indicate the pressure of metallization.

The resistivity was estimated from the measured resistance determined using Van der Pauw method. The sample thickness was estimated as 0.1 µm

# Semimetallic molecular hydrogen at pressure above 350 GPa

M. I. Eremets, A. P. Drozdov, P. P. Kong, H. Wang

Max-Planck-Institut fur Chemie, Hahn-Meitner Weg 1, 55128 Mainz, Germany

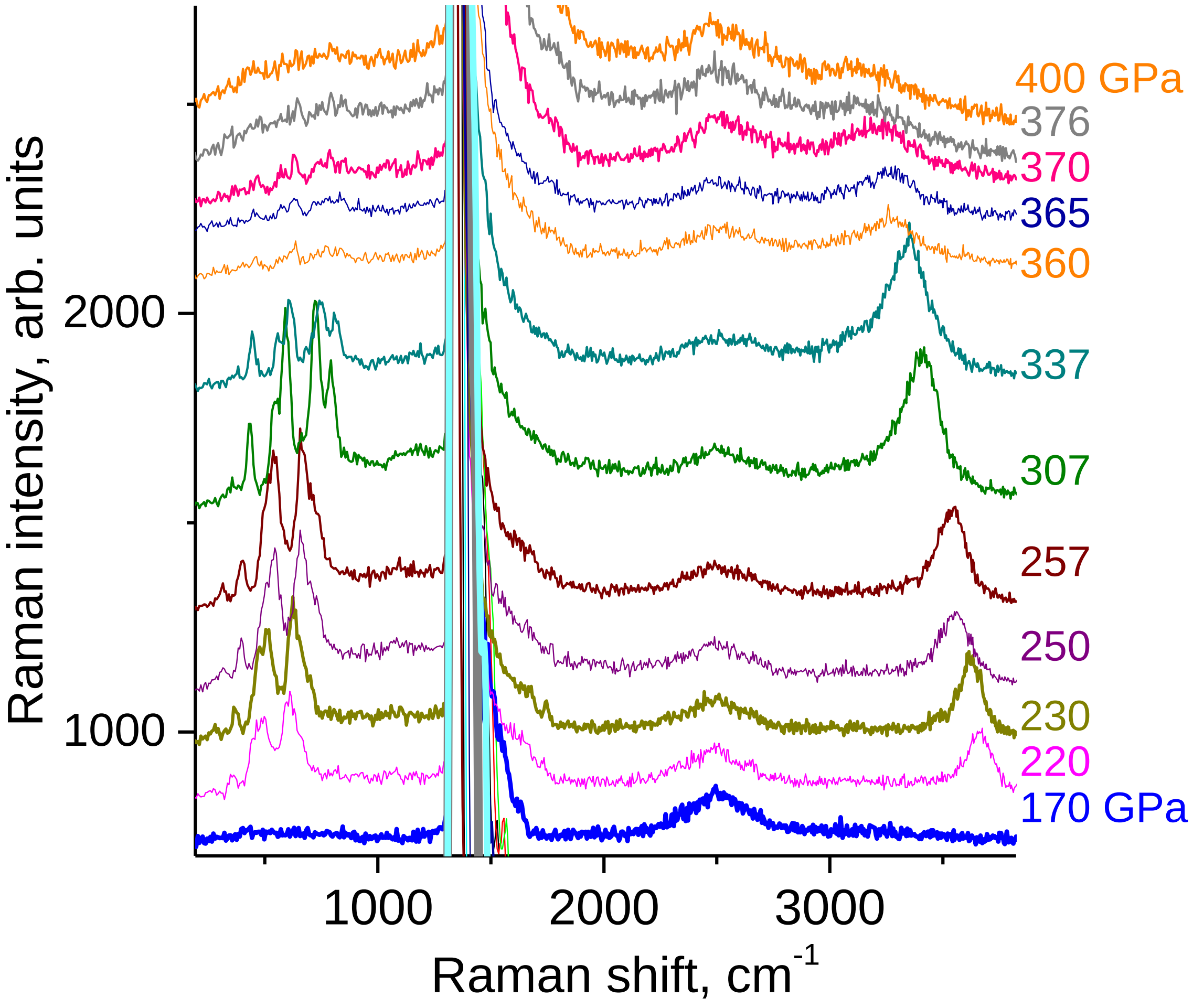

Fig. 1. Raman spectra of hydrogen taken at 100 K at different pressures.

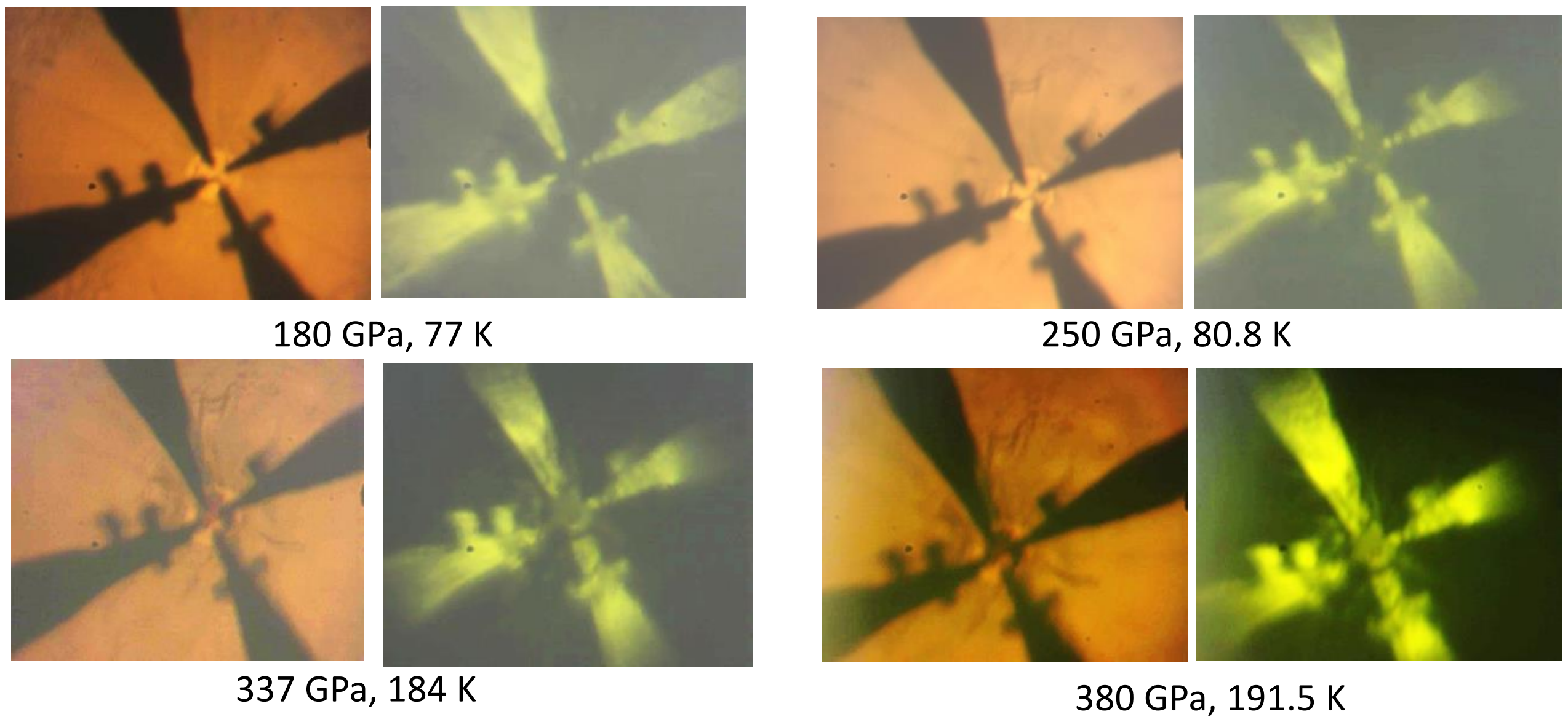

Fig. 2. Photographs of the sample of hydrogen with attached electrical leads taken at the indicated pressures and temperatures. Left pictures are taken in transmission light, right pictures – in reflection light.

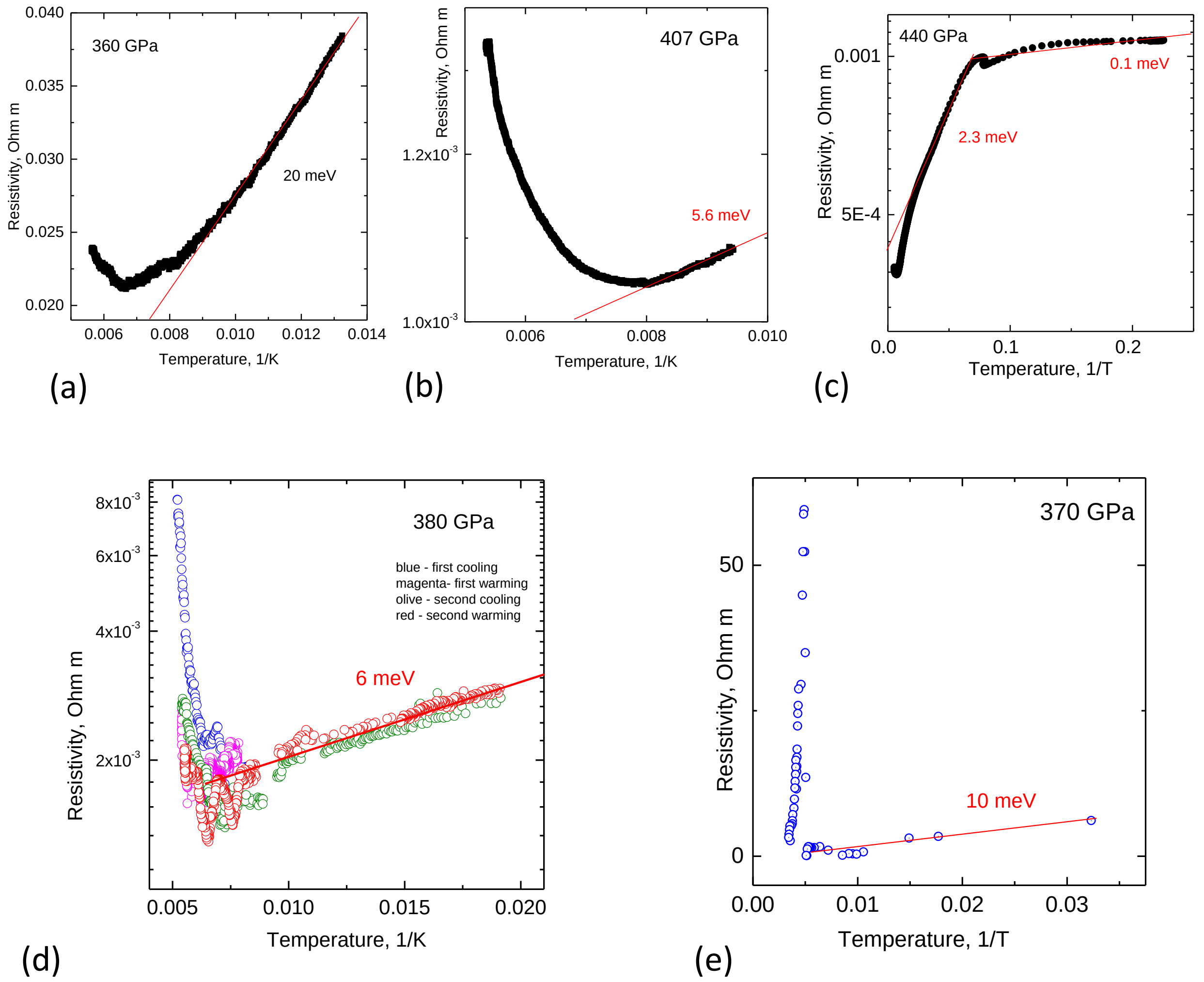

Fig. 3. The temperature dependence of resistivity $\rho(T)$ of different samples of hydrogen. $\rho(T)$ was plotted as $ln\rho$ vs $1/T$ *to* reveal characteristic activation energies. a,b,c – the graphs replotted from Fig. 3. Note that there is a metallic behavior at T>130 K which is not presentative in the reciprocal temperature scale, but it is clear seen in Fig 3b.

(c) At 440 GPa the $\rho(T)$ can be roughly approximated by two activation energies: $E_a$~6 meV at higher temperatures, and $E_a$ ~0.02 meV at the lowest temperatures where $\rho$ is nearly temperature independent.

(d) A similar $\rho(T)$ was obtained in an experiment with at 380 GPa with the $CaSO_4$ gasket. There are some oscillations in $\rho(T)$ in the 130-150 K range at some runs, but their origin is not clear.

(e) The temperature dependence of resistivity at 370 GPa taken from Ref. 13. Electrodes in this case were deposited on opposite anvils therefore a possible input to the conductivity from the diamond anvil was excluded.

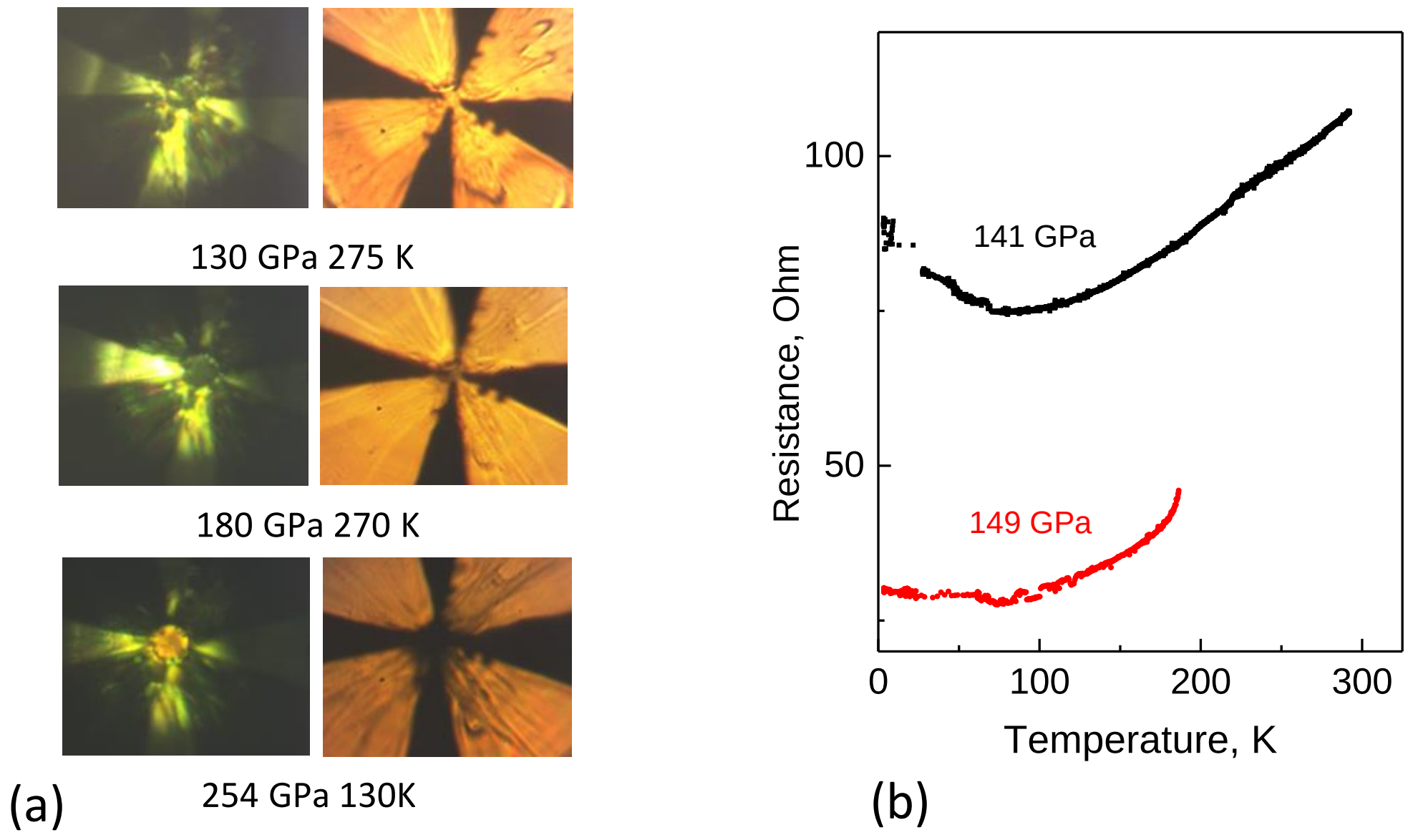

Fig. 4. Metallization of xenon. (a) Photographs of sample of xenon together with attached electrodes at the indicated pressures and temperatures. The photographs were taken in transmission or reflection modes.
(b) R(T) at the pressures close to the transition to metallic state.
The pressure dependence of the resistivity taken at ~100 K is shown in Fig. 3d (see also Ref. 26).

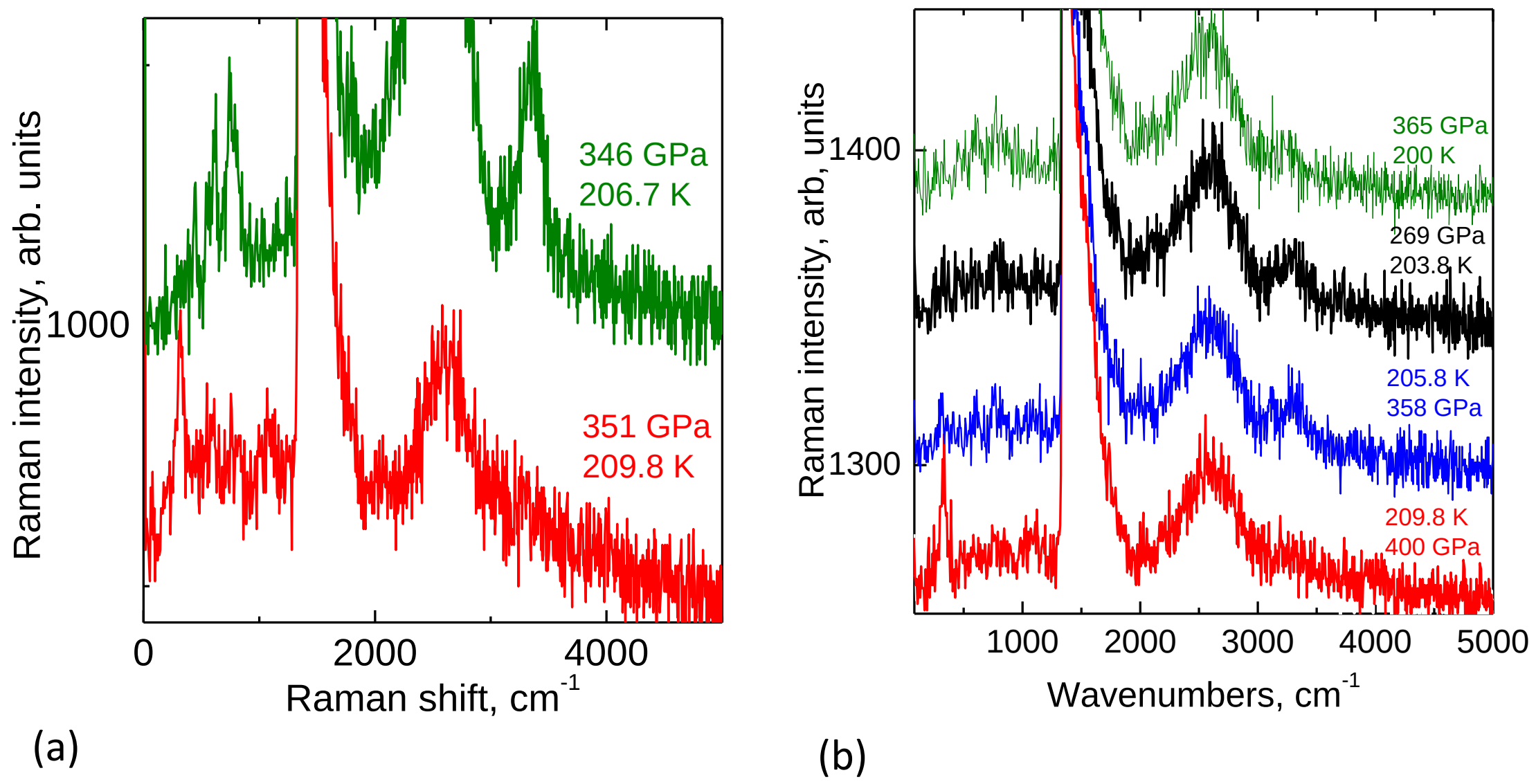

Fig. 5. Changes of the Raman spectra of hydrogen at the crossing the phase boundary between phases III and V at cooling (left panel) and warming (right panel). The phase V (red spectra) is clearly identified by the sharp peak at 330 $cm^{-1}$, also a traces of a vibron can be seen at ~4000 $cm^{-1}$, the second, broad vibron band is expected in the 2500 $cm^{-1}$ range but likely is too weak to be seen at the background of the second order scattering of diamond. At the transition, the spectra dramatically change: the features of the phase V disappear and instead the phase III appear with characteristic spectrum at low frequencies <1000 $cm^{-1}$ , and the vibron band at 3350 $cm^{-1}$ appears. Thus the temperature of the III-V transition can be located between the closest spectra at ~ 208 K at P~348 GPa and at ~ 208 K at P~370 GPa.

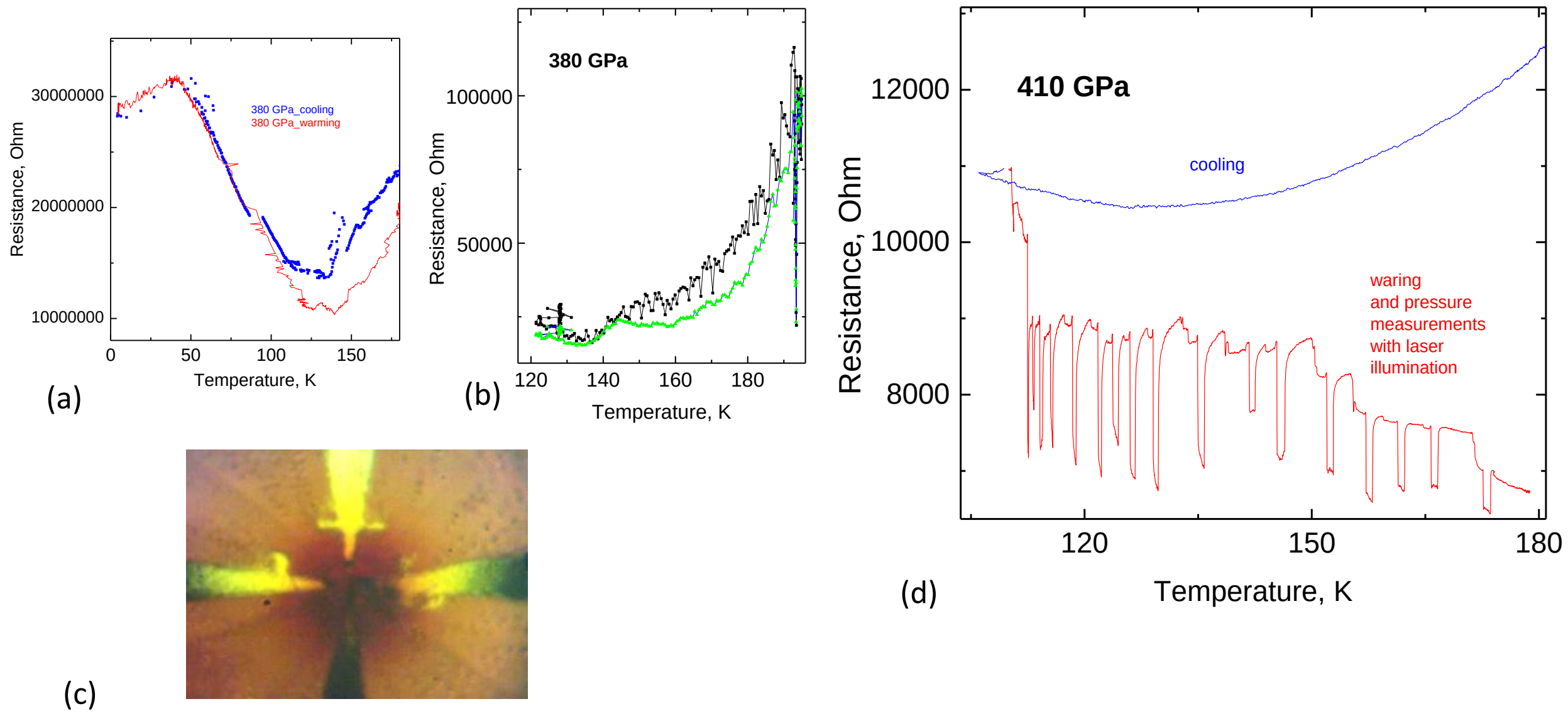

Fig. 6. Electrical measurements. (a,b) Temperature dependence of resistance measured with four electrodes:(a) and (b) are combinations of the electrodes in the van der Pauw measurements. The sample was not illuminated during the measurements. (c) hydrogen with four electrodes at >400 GPa and 45 K at the combined transmission-reflection illumination. Diamond anvils in this case are of natural Ia type. It was difficult to determine the pressure exactly because of the very high absorption of the stressed diamonds – they have brown color. d) Photoconductive response. The measured resistance is accompanied by photoconductivity at pressures P>360 GPa, i. e. it drops under the laser illumination. The upper curve (blue line) shows a measurement of the resistance upon cooling without laser illumination. Upon subsequent warming, the sample was periodically illuminated for Raman measurements. A strong drop in resistance (photoconductivity) is observed. At temperatures above 150 K, the pressure was gradually increased – this caused a decrease of resistance measured at higher temperatures. The photoconductive response is slow (minutes). We attribute this photoconductivity to diamond as no photoconductivity was observed in Ref. 12 when the electrodes were placed on the opposite anvils, and thus no conductivity from diamond was possible. In the present work, all electrodes were deposited on the anvils and the surface or bulk of conductivity or photoconductivity of diamond can interfere the measurements. Surface conductivity is unlikely as the diamonds were cleaned in argon-oxygen reactive plasma with low-energy ions. A bulk conductivity can be ruled out at first glance as diamond is a dielectric with a wide band-gap of 5.5 eV. It is well known that diamond is a good photoconductor in ultraviolet range at ambient pressure (J. Hiscock, A. T. Collins, Diamond Rel. Mater. 8, 1753 (1999). However the band gap decreases dramatically in the stressed diamond anvils and its closure is estimated to occur at 600-700 GPa (M. P. Surh et al Phys. Rev. B 45, 8239 (1992). AL Ruoff et al. J. Appl. Phys. 69, 6413 (1991). The bandgap of the diamond is indirect and thus difficult to determine from absorption measurements. However, we can determine precisely the bandgap of the stressed diamond in our experiments. An indicator is the strong luminescence, which appears at high pressures: at 360 GPa (Fig 2a) or at lower pressure in other runs: 320 GPa and 250 GPa. This luminescence was first observed by J Liu, YK Vohra, Appl. Phys. Lett. 68, 2049 (1996); we observed this luminescence in many experiments. It can arise at different pressures depending on the particular geometry of the anvils and the gasket – on particular stresses inside the anvil, and easily observed in synthetic anvils (pure material) and low temperatures. Liu and Vohra (1996) mentioned a likely electronic nature of the luminescence but did not explain it. Most likely the luminescence originates from recombination of free excitons. In this recombination process, when electronic transitions occur between the conductive and valence bands in the luminescence process, the wave vector is conserved by the emission or absorption of phonons. The strongest peak at the luminescence spectrum (Fig. 2a) is associated with the emittance of transverse-optic (TO) phonon. Therefore the band gap $E_g$ is equal to the energy of this peak plus the phonon energy. $E_{TO}$= 0.14 eV at ambient pressure, but phonons strongly stiffen with pressure so that the TO phonon frequency is equal to 0.3-0.4 eV at 400 GPa (extrapolation from K Kunc et al, Phys. Rev. B, 68, 094107 (2003)). The luminescence peak at ~2150 $cm^{-1}$ (or 1.69 eV in energy) appeared at 360 GPa (Fig.2b). Thus, the band gap is ~2.1 eV at this pressure, and decreases with pressure to ~1.65 eV at 480 GPa according to the shift of the luminescence peak. Still, optical measurements with HeNe laser ($h\omega$ = 1.959 eV) are possible because of the weak indirect absorption and very small stressed region of ~10 µm. The absorption can be very strong however in natural diamonds with impurities (c), which precludes optical measurements with the laser. The stressed diamond anvil with a band gap of 1.65-2.1 eV is an insulator, and no conductivity is expected in the 100-200 K temperature range. However, photoconductivity can occur when carriers are generated with illumination The photoconductivity should appear together with the luminescence: the generated carriers either bind to excitons and then recombine (luminescence) or participate in electrical conductivity (photoconductivity). The photoconducting response is proportional to the lifetime of the carrier, and the lifetime can be large in pure materials such as synthetic diamonds which are very pure: concentration of residual impurities is ~0.1-1 ppm.